\documentclass[longauth,a4paper]{aa}

\usepackage{graphicx}
\usepackage{txfonts}
\usepackage{natbib}
\bibpunct{(}{)}{;}{a}{}{,}

\begin{document}
  
  \title{Primary particle acceleration above 100~TeV in the shell-type
    Supernova Remnant RX~J1713.7$-$3946\ with deep H.E.S.S.\ observations}
  
  \author{F. Aharonian\inst{1}
    \and A.G.~Akhperjanian \inst{2}
    \and A.R.~Bazer-Bachi \inst{3}
    \and M.~Beilicke \inst{4}
    \and W.~Benbow \inst{1}
    \and D.~Berge \inst{1} \thanks{now at CERN, Geneva, Switzerland}
    \and K.~Bernl\"ohr \inst{1,5}
    \and C.~Boisson \inst{6}
    \and O.~Bolz \inst{1}
% ...
\email{David.Berge@cern.ch}}
  
  \institute{
    Max-Planck-Institut f\"ur Kernphysik, P.O. Box 103980, D 69029
    Heidelberg, Germany
    \and
    Yerevan Physics Institute, 2 Alikhanian Brothers St., 375036 Yerevan,
    Armenia
    \and
    Centre d'Etude Spatiale des Rayonnements, CNRS/UPS, 9 av. du Colonel Roche, BP
    4346, F-31029 Toulouse Cedex 4, France
    \and
    Universit\"at Hamburg, Institut f\"ur Experimentalphysik, Luruper Chaussee
    149, D 22761 Hamburg, Germany
    \and
    Institut f\"ur Physik, Humboldt-Universit\"at zu Berlin, Newtonstr. 15,
    D 12489 Berlin, Germany
    \and
    LUTH, UMR 8102 du CNRS, Observatoire de Paris, Section de Meudon, F-92195 Meudon Cedex,
    France
% ...
}
  
  \date{Received 11 September 2006 / Accepted 22 November 2006}
  
  \abstract
      {}
      {We present deep H.E.S.S.\ observations of the supernova
	remnant (SNR) RX~J1713.7$-$3946. Combining data of three years -- from 2003
	to 2005 -- we obtain significantly increased statistics
	and energy coverage as compared to earlier 2003 \& 2004
	results.}
      {The data are analysed separately for the different years.}
      {Very good agreement of the gamma-ray morphology and the
	differential spectra is found when comparing the three
	years. The combined gamma-ray image of the 2004 and 2005 data
	reveals the morphology of RX~J1713.7$-$3946\ with unprecedented
	precision. An angular resolution of $0.06\degr$ is achieved,
	revealing the detailed structure of the remnant. The combined
	spectrum of all three years extends over three orders of
	magnitude, with significant gamma-ray emission approaching
	100~TeV. The cumulative significance above 30~TeV is
	$4.8\sigma$, while for energies between 113~and~294~TeV an upper
	limit on the gamma-ray flux of $1.6\times
	10^{-16}~\mathrm{cm}^{-2}~\mathrm{s}^{-1}$ is obtained.}
      {The energy coverage of the H.E.S.S.\ data is presumably at the
	limit of present generation Cherenkov telescopes. The
	measurement of significant gamma-ray emission beyond 30~TeV
	formally implies the existence of primary particles of at
	least that energy. However, for realistic scenarios of
	very-high-energy gamma-ray production, the Inverse Compton
	scattering of very-high-energy electrons and $\pi^0$ decay
	following inelastic proton-proton interactions, the measured
	gamma-ray energies imply that efficient acceleration of
	primary particles to energies exceeding 100~TeV is taking
	place in the shell of the SNR RX~J1713.7$-$3946.}

  \authorrunning{Aharonian et al.}
  \titlerunning{Deep H.E.S.S.\ observations of RX~J1713.7$-$3946}
  
  \keywords{acceleration of particles -- ISM: cosmic rays -- gamma rays:
    observations -- ISM: supernova remnants -- gamma rays:
    individual objects: \object{RX~J1713.7$-$3946}
    (\object{G347.3$-$0.5})} 
  
  \maketitle

  \section{Introduction}
  \label{sec:intro}

  \begin{table*}
    \label{table:obstimes}
    \centering
    \begin{tabular}{|c||cc|cc|c|c|c|c|}
      \hline 
      & \multicolumn{2}{c|}{Target position} & \multicolumn{2}{c|}{Wobble
	offsets} & Live time & Data set I & Data set II & Data set III\\
      Year & & & & & & \emph{best-resolution data} & \emph{spectral
      comparisons} & \emph{all data}\\
      & $\alpha_{\mathrm{J}2000}$ 
      &
      $\delta_{\mathrm{J}2000}$ & $\Delta \alpha_{\mathrm{J}2000}$ & $\Delta
      \delta_{\mathrm{J}2000}$ & (hours) & (hours) & (hours) & (hours)\\\hline\hline
      & & & $0\degr$ & $+0.5\degr$ & 9.1 & & &\\
      \raisebox{1.5ex}[-1.5ex]{2003} & \raisebox{1.5ex}[-1.5ex]{17h12m00s} & \raisebox{1.5ex}[-1.5ex]{$-39\degr 39\arcmin 00\arcsec$} &
      $0\degr$ & $-0.5\degr$ & 9.0 & \raisebox{1.5ex}[-1.5ex]{0} & \raisebox{1.5ex}[-1.5ex]{18.1} & \raisebox{1.5ex}[-1.5ex]{18.1}\\\hline
      & & & $0\degr$ & $+0.7\degr$ & 7.3 & & &\\
      & & & $0\degr$ & $-0.7\degr$ & 8.1 & & &\\
      2004 & 17h13m33s & $-39\degr45\arcmin 44\arcsec$ & $+0.7\degr$ &
      $0\degr$ & 6.9 & 33.1 & 30.6 & 33.1\\
      & & & $-0.7\degr$ & $0\degr$ & 8.3 & & &\\
      & & & $0\degr$ & $0\degr$ & 2.5 & & &\\\hline
      & & & $0\degr$ & $+0.7\degr$ & 9.0 & & &\\
      & & & $0\degr$ & $-0.7\degr$ & 9.5 & & &\\
      2005 & 17h13m33s & $-39\degr45\arcmin 44\arcsec$ & $+0.7\degr$ &
      $0\degr$ & 8.7 & 29.7 & 36.8 & 40.1\\
      & & & $-0.7\degr$ & $0\degr$ & 9.4 & & & \\
      & & & $0\degr$ & $0\degr$ & 3.5\footnotemark[1] & & &\\\hline
      \multicolumn{9}{l}{
	\begin{minipage}[t]{0.9\textwidth}
	  \footnotemark[1] These observations had wobble offsets of
	  $0.5\degr$ in Right Ascension and Declination, less than the
	  radius of the SNR, and are summarised in one row. 
	\end{minipage}
      }
    \end{tabular}\\[1.0ex]
    \caption{Summary of H.E.S.S.\ observations of RX~J1713.7$-$3946\ conducted during
      three years. For each year, the targeted position is given
      together with the wobble offsets in Right Ascension
      $(\alpha_\mathrm{J2000})$ and Declination
      $(\delta_\mathrm{J2000})$. Adding the wobble offsets to the target
      coordinates, the actual pointing position is obtained. For each
      pointing position, the dead-time corrected observation time
      (\emph{Live time}) is given in hours. Data recorded under bad
      weather conditions are excluded. The columns \emph{Data set
      I-III} summarise observation times of data sub-sets used
      throughout the paper. \emph{Data set I} is used for
      Fig.~\protect\ref{fig:CombinedImage}; to obtain optimum angular
      resolution, the 2003 data are disregarded and a zenith-angle cut
      at $60\degr$ is applied (the latter is only relevant for the
      2005 data). \emph{Data set II} is used for spectral comparisons
      of the different years (cf.\
      Fig.\protect\ref{fig:SpectraAllYears}). Observations with wobble
      offsets of $0\degr$ in 2004 and 2005 are disregarded for this
      purpose. \emph{Data set III} comprises all data, and is used for
      the generation of the combined spectrum (cf.\
      Fig.~\protect\ref{fig:CombinedSpectrum}).}
  \end{table*}

  The energy spectrum of cosmic rays measured at Earth exhibits a
  power-law dependence over a broad energy range. Starting at a few
  GeV $(1~\mathrm{GeV} = 10^9~\mathrm{eV})$ it continues to energies of at
  least $10^{20}~\mathrm{eV}$. The power-law index of the spectrum
  changes at two characteristics energies: in the region around $3
  \times 10^{15}~\mathrm{eV}$ -- the \emph{knee} region -- the spectrum
  steepens, and at energies beyond $10^{18}~\mathrm{eV}$ it hardens
  again. This latter feature is known as the \emph{ankle}. Up to the
  knee, cosmic rays are believed to be of Galactic origin, accelerated
  in shell-type supernova remnants (SNRs) -- expanding shock waves
  initiated by supernova explosions~(for a recent review see
  \citet{HillasReview}). However, the experimental confirmation of an
  SNR origin of Galactic cosmic rays is difficult due to the
  propagation effects of charged particles in the interstellar
  medium. The most promising way of proving the existence of
  high-energy particles in SNR shells is the detection of
  very-high-energy (VHE) gamma rays ($E > 100~\mathrm{GeV}$), produced in
  interactions of cosmic rays close to their acceleration
  site~\citep{DAV}.

  Recently H.E.S.S.~--~a VHE gamma-ray instrument consisting of four
  Imaging Atmospheric Cherenkov Telescopes~--~has detected two
  shell-type SNRs, RX~J1713.7$-$3946~\citep{Hess1713a,Hess1713b} and
  RX~J0852.0--4622~\citep{HessVelaJr_a}. The two objects show an
  extended morphology and exhibit a shell structure, as expected from
  the notion of particle acceleration in the expanding shock
  fronts. Both objects reveal gamma-ray spectra that can be described
  by a hard power law (with photon index $\Gamma \sim 2.0$) over a
  broad energy range. For RX~J1713.7$-$3946\ significant deviations from a pure
  power law at larger energies are measured~\citep{Hess1713b}. While
  it is difficult to attribute the measured VHE gamma rays
  unambiguously to nucleonic cosmic rays (rather than to cosmic
  electrons, which would certainly also be accelerated in the shock
  front), the measured spectral shapes favour indeed in both cases --
  for RX~J1713.7$-$3946\ and RX~J0852.0--4622 -- a nucleonic cosmic-ray origin of
  the gamma rays~\citep{HessVelaJr_b,Hess1713b}. In the case of RX~J1713.7$-$3946\
  in addition a narrow shock filament seen in X-rays~\citep{HiragaXMM}
  indicates strong amplification of the magnetic field at least in one
  region of the rim~\citep{BerezhkoVoelk}. If such an amplified
  magnetic field exists throughout the main volume of the SNR~--~the
  region for which VHE gamma-ray data is presented here~--~and if
  consequently high magnetic field values are found not only in one
  shock filament, but on a large part of the shock surface, a leptonic
  origin of the VHE gamma rays becomes increasingly unlikely just
  based on the absolute level of X-ray and gamma-ray flux of
  RX~J1713.7$-$3946~\citep{Hess1713b}.

  Apart from the first unambiguous proof of multi-TeV particle
  acceleration in SNRs~\citep{HessVelaJr_b,Hess1713b}, the question of
  the highest observed energies remains an important one. Only the
  detection of gamma rays with energies of 100~TeV and beyond provides
  experimental proof of acceleration of primary particles, protons or
  electrons, to even higher energies of 1~PeV and beyond. The spectrum
  of the whole SNR RX~J1713.7$-$3946\ reported in~\citet{Hess1713b} comprises data
  of the 2004 observation campaign of H.E.S.S.. It extends to energies of
  40~TeV. Here we present a combined analysis of H.E.S.S.\ data of RX~J1713.7$-$3946\
  recorded in three years, in 2003 during the construction and
  commissioning phase of the system, and in 2004 and 2005 with the
  full H.E.S.S.\ array. A comparison of the three data sets demonstrates
  the expected steady emission of the source as well as the stability
  of the system during the first three years of running. Special
  emphasis is then devoted to the high-energy end of the combined
  spectrum.

  \section{H.E.S.S.\ observations}
  \label{sec:hess_obs}

  The High Energy Stereoscopic System (H.E.S.S.) consists of four
  identical Cherenkov telescopes that are operated in the Khomas
  Highland of Namibia~\citep{HofmannStatusHess}. The
  telescopes~\citep{Konrad2003,CornilsII} are 13~m in diameter, each
  with a mirror area of $107~\mathrm{m}^{2}$. During normal operation the
  system is run in a coincidence mode which requires a trigger from at
  least two out of the four telescopes~\citep{FunkTriggerPaper}. The
  cameras consist of 960 photomultiplier pixels and cover a $5\degr$
  field of view~\citep{PascalHessCamera}. The resulting $FWHM \approx
  4\degr$ of the system field-of-view response makes H.E.S.S.\ the
  currently best suited experiment in the field for the study of
  extended VHE gamma-ray sources such as young Galactic SNRs. At
  zenith, the energy threshold is about 100~GeV and for point sources
  an energy resolution of 15\% is achieved. The angular resolution for
  individual gamma rays is better than $0.1\degr$ and the point source
  sensitivity reaches $1\%$ of the flux of the Crab nebula for long
  exposures ($\sim25$~hours).

  The H.E.S.S.\ observation campaign of RX~J1713.7$-$3946\ started in 2003. The data
  were recorded between May and August 2003 during two phases of the
  commissioning of the telescope system. During the first phase, two
  telescopes were operated independently with stereoscopic event
  selection done offline using GPS time stamps to identify coincident
  events. During the second phase, also using two telescopes,
  coincident events were selected in hardware using the array level
  trigger~\citep{FunkTriggerPaper}. The observations were performed in
  Declination wobble mode around the northwest shell of the SNR, the
  alternating wobble offset in Declination was $0.5\degr$. The zenith
  angle of observations varied from $15\degr$ to $30\degr$ with a mean
  of $24\degr$. The analysis of this first data set revealed extended
  gamma-ray emission resembling a shell structure, very similar to the
  X-ray image. It was actually the first ever resolved image of an
  astronomical source obtained with VHE gamma
  rays~\citep{Hess1713a}. The spectrum was well described by a hard
  power law with energies from 1 to 10~TeV.
    
  In 2004, observations were conducted with the full telescope
  array. From April to May, most of the data were recorded in wobble
  mode, this time around the SNR centre with an offset of $0.7\degr$
  in Right Ascension and Declination aiming at more uniform coverage
  of the whole SNR and, important for analysis purposes, fully
  encompassing the SNR with the four observation positions. The zenith
  angle of observations ranged from $16\degr$~to~$56\degr$ with a mean
  of $26\degr$. The H.E.S.S.\ data enabled analysis of the gamma-ray
  morphology and the spectrum of the remnant with unprecedented
  precision~\citep{Hess1713b}. A very good correlation was found
  between the X-ray and the gamma-ray image. The differential spectrum
  was measured from 200~GeV up to 40~TeV. A deviation from a pure
  power law was found at high energies. A spatially resolved spectral
  study revealed no significant changes of spectral shape across the
  SNR despite flux variations by more than a factor of two.

  \begin{table*}
    \label{table:hard_cuts_stats}
    \centering
    \begin{tabular*}{\textwidth}{@{\extracolsep{\fill}}|c||*{7}{c|}}
      \hline 
      Year & $\langle \phi_\mathrm{z} \rangle$ & $R_{68}$ & ON & OFF &
      $\alpha$ & Significance $(\sigma)$ & Live time
      (hrs)\\\hline\hline
      2003 & $24\degr$ & $0.083\degr$ &
      3194 & 1764 & 1.00 & 21 & 18.1\\\hline
      2004 & $27\degr$ & $0.075\degr$\ & 18728
      & 11039 & 1.05 & 41 & 33.1\\\hline
      2005 & $44\degr$ & $0.082\degr$\ & 10277
      & 5124 & 1.15 & 33 & 29.7\\\hline
    \end{tabular*}
    \caption{Summarised are the event statistics of the whole SNR and
      corresponding angular resolutions for the years 2003, 2004, and
      2005. The 2004 and 2005 data sets correspond to \emph{Data set
      I} of Table~\ref{table:obstimes}. The average zenith angle
      $\langle \phi_\mathrm{z} \rangle$ is determined from all events
      reconstructed in the SNR region. For the angular resolution
      ($R_{68}$), the 68\% containment radius of the
      simulated gamma-ray point-spread function, matched to the
      particular data set, is used as figure of merit. The
      other columns give the number of signal events from the SNR
      region (ON), the number of background events (OFF), the
      normalisation factor between ON and OFF counts ($\alpha$), and
      the corresponding significance and live observation
      time. $\alpha$ is in general defined as the ratio of the
      effective exposure integrated in time and angular space of the
      ON and OFF region. Note that the analysis of the 2003 data is
      adopted to match the system configuration of this year. The
      nominal analysis is applied for the 2004 and 2005 data. For
      2005, only data recorded at zenith angles less than $60\degr$
      are included (therefore the mean zenith angle decreases).  The
      event statistics are determined with the \emph{ON/OFF} approach
      for 2003 and the \emph{reflected-region} method for 2004 and
      2005. In the latter two years, also \emph{ON} runs with wobble
      offsets $<0.7\degr$~(cf.\ Table~\ref{table:obstimes}) are
      included and hence $\alpha>1$.}
  \end{table*}

  The 2005 observation campaign was aiming at extending the energy
  coverage of the spectrum to as high energies as possible. Therefore
  the observations, carried out from beginning of September to
  November, were preferentially pursued at large zenith angles, up to
  values of $70\degr$, to make use of the drastically increased
  effective collection area of the experiment at high energies. The
  mean zenith angle of observations was $51\degr$. As in 2004, RX~J1713.7$-$3946\
  was observed in wobble mode with an offset of $0.7\degr$ in
  Declination and Right Ascension. Analysis of these data are for the
  first time presented in the following. A summary of the observations
  conducted during three years with H.E.S.S.\ is given in
  Table~\ref{table:obstimes}.
  
  \section{Data Analysis}
  \label{sec:data_analysis}

  The RX~J1713.7$-$3946\ data presented here are calibrated according to the
  standard H.E.S.S.\ calibration methods~\citep{HessCalib}. For the
  background suppression, cuts on scaled image parameters are
  applied~\citep{Hess2155}. The shower reconstruction is based on
  image parameters (\emph{Hillas} parameters) and corresponds, unless
  otherwise stated, to \emph{algorithm~1} of
  \citet{HofmannShowerReco}: the intersection point of the image axes
  in a common camera coordinate system yields the shower impact
  position on ground and the direction of the primary. A cut on the
  minimum size of camera images is applied to assure that only well
  defined images are included in the analysis. For the 2003
  two-telescope data, the cut is applied at a rather large value of
  300 photo-electrons. In the commissioning phase of the experiment,
  this served to dramatically reduce the number of background events,
  but it also homogenises the whole data set, which was recorded with
  two different hardware configurations, thereby reducing systematic
  uncertainties. Moreover, the angular resolution improves when
  including only well defined images in the analysis. The 2004 and
  2005 data are analysed as discussed in \citet{Hess1713b}. For
  spectral analysis, a loose cut on the minimum image size at 80
  photo-electrons is applied. For studies of the gamma-ray morphology,
  the cut is increased to 200 photo-electrons yielding superior
  angular resolution of the order of $0.08\degr$ and better background
  suppression.

  For the subtraction of the irreducible cosmic-ray background,
  separate approaches are taken for the generation of gamma-ray
  spectra and images. The preferred background-estimation method for
  spectral analysis is the \emph{reflected-region}
  model~\citep{BackgroundProc}. The background estimate is derived
  from a region of the same size and shape as the source region,
  reflected at the system pointing direction. To assure
  non-overlapping source and background-control regions, this approach
  can only be applied if the observation positions have been chosen
  outside the nominal gamma-ray source region. As can be seen from
  Table~\ref{table:obstimes}, this is not true for the whole 2003 data
  and parts of the 2004 and 2005 observations. For these data, an
  \emph{ON/OFF}-background model is applied instead. From the complete
  set of H.E.S.S.\ observations without gamma-ray signal, \emph{OFF} runs
  for background estimation are selected with zenith-angle
  distributions matching that of the \emph{ON} runs as close as
  possible.

  For image generation, the \emph{field-of-view}-background model is
  applied~\citep{BackgroundProc}. It models the background by means of
  a system acceptance model determined from the full set of H.E.S.S.\
  \emph{OFF} runs. The normalisation is calculated using the whole
  field of view excluding regions of known gamma-ray emission. Note
  that the background-subtracted gamma-ray images shown throughout
  this paper are smoothed with a Gaussian to reduce statistical
  fluctuations. The resulting images are in units of gamma-ray excess
  counts per Gaussian sigma. They are corrected for the falloff of the
  system acceptance towards the edges of the field of view which
  results from a smaller detection efficiency far from the pointing
  centre.

  When determining spectra of the whole SNR, a circular region of
  $0.65\degr$ radius is used here, centred at $\alpha_{\mathrm{J2000}} \,
  = \, \mathrm{17h13m33.6s}$, $\delta_{\mathrm{J2000}} \, = \, -39\degr
  45\arcmin 36\arcsec$.

  \begin{figure*}
    \centering
    \includegraphics[width=17cm,draft=false]{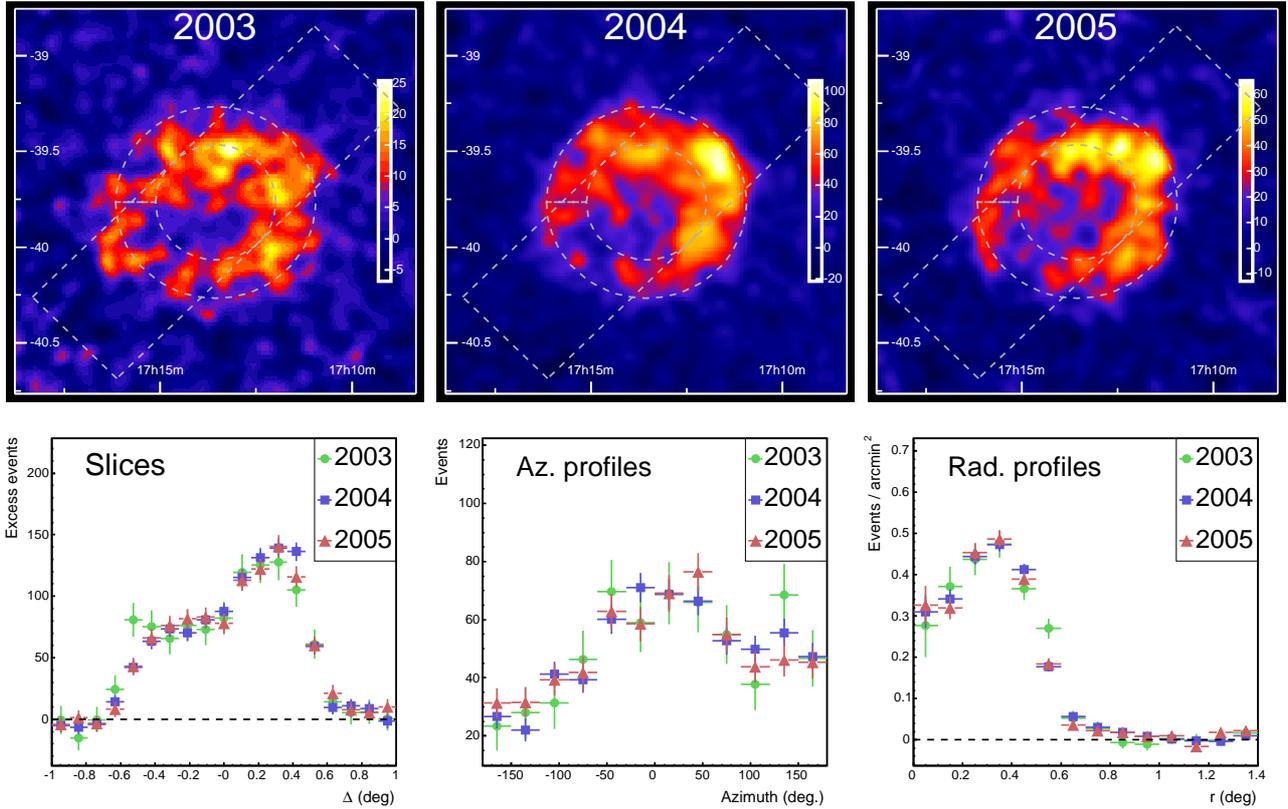}
    \caption{\underline{\textbf{Upper panel:}} H.E.S.S.\ gamma-ray excess
      images from the region around RX~J1713.7$-$3946\ are shown for each year
      separately for comparison. From left to right, images are
      generated from data of 2003, 2004 and 2005. The images are
      corrected for the decline of the system acceptance with
      increasing distance to the SNR centre. All three images are
      smoothed with a Gaussian of $2\arcmin$, the linear colour scale
      is in units of excess counts per smoothing radius. The dashed
      box (dimensions $2\degr \times 0.6\degr$) and ring ($r_1 =
      0.3\degr$, $r_2 = 0.5\degr$) are used for obtaining the
      one-dimensional distributions shown in the lower
      panel. \underline{\textbf{Lower panel:}} One-dimensional
      distributions for the data of three years, all generated from
      the raw, non-smoothed and acceptance-corrected gamma-ray excess
      images. Histograms of 2004 and 2005 are scaled to the area of
      the 2003 histogram to account for differences in the event
      statistics. \textbf{Left:} Slices taken within a rotated box
      running through the SNR region. Plotted are events versus
      angular distance to the centre, projected onto the axis running
      through the SNR centre, rotated by $45\degr$ anti-clockwise with
      respect to the RA axis. \textbf{Middle:} Azimuth profiles
      integrated in a thick ring covering the shell of RX~J1713.7$-$3946. The
      azimuthal angle of the events is calculated with respect to the
      SNR centre ($\alpha_{\mathrm{J2000}} \, = \, \mathrm{17h13m33.6s}$,
      $\delta_{\mathrm{J2000}} \, = \, -39\degr 45\arcmin
      36\arcsec$). $0\degr$ corresponds to the west part of the shell,
      $90\degr$ is north or upward, $-90\degr$ is south or
      downward. \textbf{Right:} Radial profiles around the centre of
      the SNR. Plotted are excess events per unit solid angle as a
      function of the distance $r$ to the SNR centre.}
    \label{fig:ImagesAllYears}
  \end{figure*}

  \section{Gamma-ray Morphology}
  \label{sec:gamma_ray_morph}
  
  \begin{figure*}
    \centering
    \includegraphics[width=17cm,draft=false]{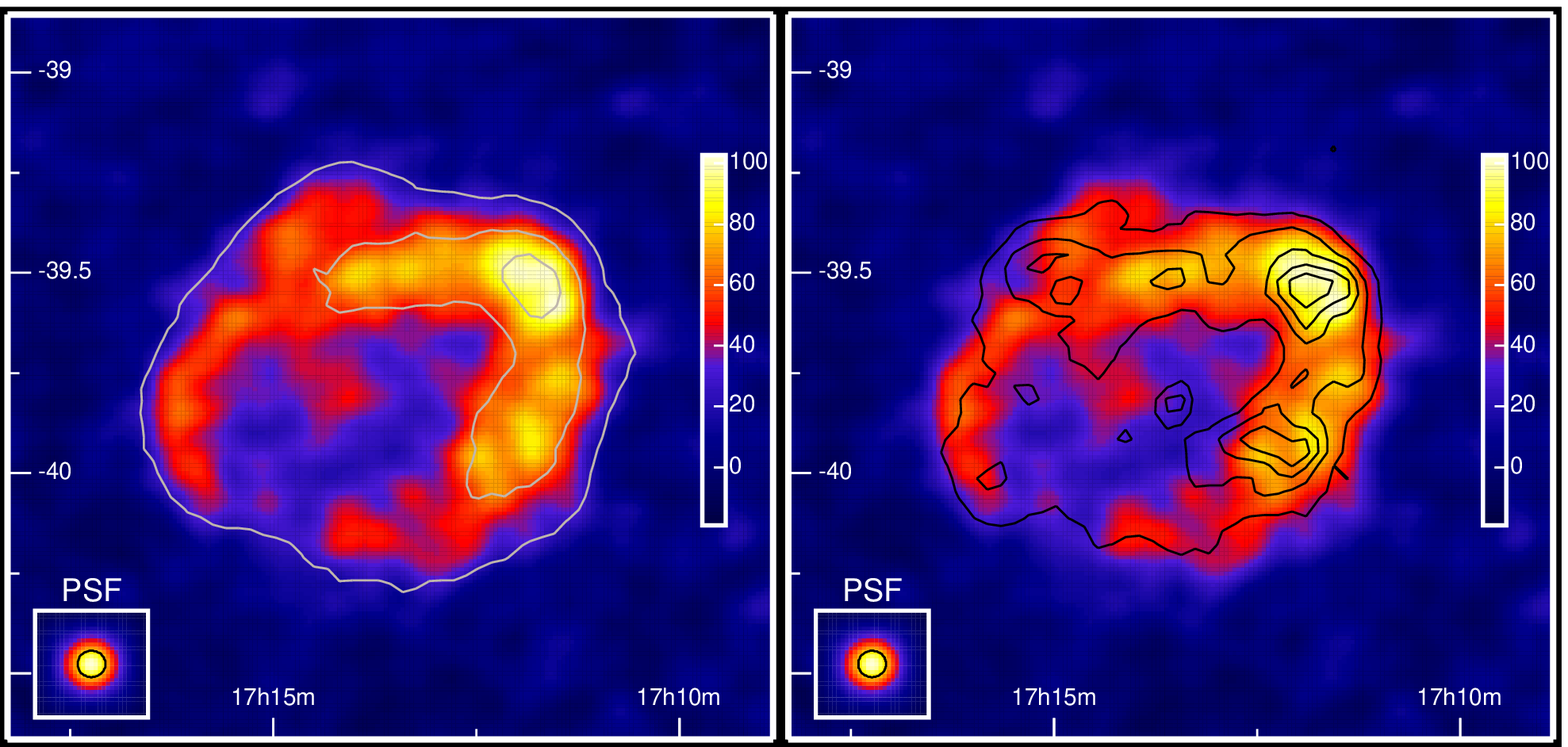}
    \caption{Two versions of the combined H.E.S.S.\ image from the 2004
      and 2005 data. Shown is in both cases an acceptance-corrected
      gamma-ray excess image. The images are smoothed with a Gaussian
      of $2\arcmin$. A simulated point source as it would appear in
      this data set is shown in the lower left-hand corner of both
      images (labeled \emph{PSF}). It is smoothed with the same
      Gaussian, the $\sigma$ of $2\arcmin$ is denoted as black circle
      in the insets. The linear colour scale is in units of excess
      counts per smoothing radius. Note that for the 2005 data, only
      data recorded at zenith angles less than $60\degr$ are taken
      into account. On the \textbf{left-hand side}, the overlaid
      light-gray contours illustrate the significance of the different
      features. The levels are at $8$, $18$, and $24\sigma$. The
      significance at each position has been calculated for a
      point-source scenario, integrating events in a circle of
      $0.1\degr$ radius around that position. On the
      \textbf{right-hand side}, ASCA contours are drawn as black
      lines~(1-3~keV, from \citet{UchiyamaAsca}) for comparison.}
    \label{fig:CombinedImage}
  \end{figure*}

  The gamma-ray morphology as obtained from the H.E.S.S.\ data of three
  years is seen in the upper panel of
  Fig.~\ref{fig:ImagesAllYears}. For the 2003 data, a special set of
  analysis cuts was applied adopted to the two-telescope data~(cf.\
  Sec.~\ref{sec:data_analysis}). For the 2005 image, only observations
  at zenith angles smaller than $60\degr$ are taken into account
  (reducing the available observation time by $\sim10$~hours, cf.\
  Table~\ref{table:obstimes}). For larger zenith-angle observations,
  the geometrical reconstruction worsens, deteriorating the angular
  resolution of the resulting image.

  The images of Fig.~\ref{fig:ImagesAllYears} are readily comparable.
  They are corrected for system acceptance, which is different for the
  different data because of the zenith-angle dependence of the
  acceptance and the intermediate system configuration of 2003. Very
  similar angular resolutions are achieved for all years, see
  Table~\ref{table:hard_cuts_stats}, where relevant parameters are
  listed. From the visual impression the three images shown in the
  figure are very similar. Within statistics, good agreement is
  achieved, as can be seen from the one-dimensional distributions
  shown in the lower panel of Fig.~\ref{fig:ImagesAllYears}, which
  have the advantage that statistical errors on the measurement can be
  taken into account for the comparison. Shown from left to right are
  a slice along a thick box (shown in the upper panel of
  Fig.~\ref{fig:ImagesAllYears}), an azimuthal profile of the shell
  region, and a radial profile. All the distributions are generated
  from the non-smoothed, acceptance-corrected excess images, very
  finely binned such that binning effects are negligible. Clearly,
  there is no sign of disagreement or variability, the H.E.S.S.\ data of
  three years are well compatible with each other.
  
  The combined H.E.S.S.\ image is shown in
  Fig.~\ref{fig:CombinedImage}. Data of 2004 and 2005 are used for
  this smoothed, acceptance-corrected gamma-ray excess image
  (\emph{Data set I} in Table~\ref{table:obstimes}). In order to
  obtain optimum angular resolution, a special analysis is applied
  here. In addition to the image-size cut of 200~photo-electrons, the
  cut on the minimum event multiplicity is raised to three telescopes
  (disregarding the 2003 data for this purpose completely). Moreover,
  an advanced reconstruction method is chosen. It takes Monte-Carlo
  error estimates on image parameters into account and is based on
  \emph{algorithm~3} of \citet{HofmannShowerReco}~(see
  \citet{PhDBerge} for studies of this analysis technique). The image
  corresponds to 62.7 hours of dead-time corrected observation
  time. With the \emph{reflected-region} method, 12961 ON events from
  the region associated with the SNR are accumulated, and 5710 OFF
  events (normalisation $\alpha = 1.1$\footnote{Note that pure
  \emph{ON} runs with wobble offsets $<0.7\degr$ are included in
  \emph{Data set I}~(cf.\ Table~\ref{table:obstimes}) and hence
  $\alpha>1$.}). Hence, 6702 gamma-ray excess events are measured
  with a statistical significance of $48\sigma$. An angular resolution
  of $0.06\degr$ ($3.6\arcmin$) is achieved. For comparison, the
  resolution obtained with the standard geometrical reconstruction
  method and a three-telescope multiplicity is $0.07\degr$ with
  similar event statistics. With a two-telescope multiplicity cut, the
  resolution with the standard reconstruction is $0.08\degr$ (with
  28879 ON, 16070 OFF events, $\alpha = 1.1$, and a significance of
  53~$\sigma$).

  The image in Fig.~\ref{fig:CombinedImage} confirms nicely the
  published H.E.S.S.\ measurements~\citep{Hess1713a,Hess1713b}, with 20\%
  better angular resolution and increased statistics. The shell of
  RX~J1713.7$-$3946, somewhat thick and asymmetric, is clearly visible and almost
  closed. As can be seen from the left-hand side of the figure, when
  integrating signal and background events in a circle of $0.1\degr$
  radius around each trial point-source position, significant
  gamma-ray emission is found throughout the whole remnant. Even in
  the seemingly void south-eastern region it exceeds a level of $8$
  standard deviations. The gamma-ray brightest parts are located in
  the north and west of the SNR. The similarity of gamma-ray and X-ray
  morphology, which was already investigated in detail
  in~\citet{Hess1713b} for the 2004 H.E.S.S.\ data, is again demonstrated
  on the right-hand side of Fig.~\ref{fig:CombinedImage}, where ASCA
  X-ray contours are overlaid on the H.E.S.S.\ image.

  \section{Gamma-ray Spectrum}
  \label{sec:gamma_ray_spectrum}
  
  \begin{table*}
    \centering
    \begin{tabular*}{\textwidth}[hbtp]{@{\extracolsep{\fill}}|c||*{5}{c|}}
      \hline Year & ON & OFF & $\alpha$ & Significance $(\sigma)$ &
      Live time (hrs)\\\hline\hline %
      2003 & 3194 & 1764 & 1.0 & 20.5 & 18.1\\\hline
      2004 & 107494 & 93906 & 1.0 & 30.3 & 30.6\\\hline
      2005 & 71276 & 60175 & 1.0 & 30.6 & 36.8\\\hline
    \end{tabular*}
    \caption{Comparison of event statistics from the SNR region from
      three years of data. The numbers result from the spectral
      analysis of \emph{Data set II} (cf.\
      Table~\ref{table:obstimes}), shown in
      Fig.~\ref{fig:SpectraAllYears}. Given are the number of signal
      (\emph{ON}) and background (\emph{OFF}) counts, the
      normalisation factor $\alpha$, the statistical significance of
      the gamma-ray excess ($\sigma$) and the observation time. For
      the 2003 data, the special two-telescope analysis with a cut on
      the minimum size of camera images at 300 photo-electrons was
      applied. The background estimate in this case is derived with
      the \emph{ON/OFF} analysis. For 2004 and 2005, the nominal
      spectral analysis with a cut at 80 photo-electrons was used
      together with \emph{reflected-region} background model.}
    \label{table:std_cuts_stats}
  \end{table*}

  \begin{figure}
    \resizebox{\hsize}{!}{\includegraphics[draft=false]{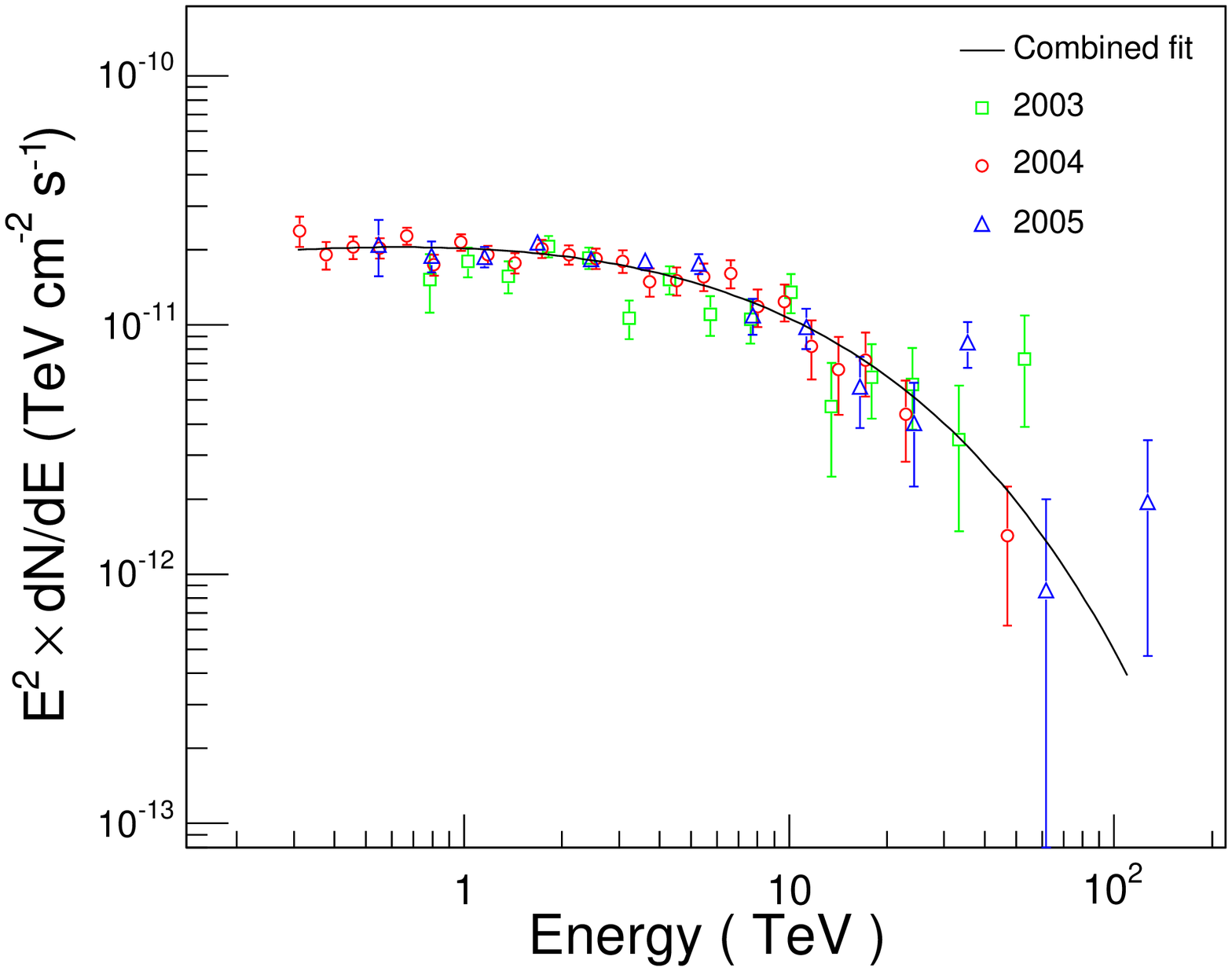}}
    \caption{Comparison of H.E.S.S.\ spectra from the years 2003, 2004,
      and 2005 (\emph{Data set II}, Table~\ref{table:obstimes}). The
      three spectra are shown in an energy-flux representation -- flux
      points have been multiplied by $E^2$. The black curve is shown
      for reference. It is the best fit of a power law with
      exponential cutoff to the combined data, where the cutoff is
      taken to the power of $\beta = 0.5$: $dN / dE = I_0 \,
      E^{-\Gamma} \,
      \exp\left(-(E/E_\mathrm{c})^{~\beta=0.5}\right)$. Note that flux
      points are corrected for the degradation of the optical
      efficiency of the system. The energy threshold of $\sim1$~TeV in
      the 2003 data is due to the two-telescope operation mode and the
      application of a stringent cut on the minimum camera image
      size.}
    \label{fig:SpectraAllYears}
  \end{figure}

  The gamma-ray spectra measured with H.E.S.S.\ in three consecutive
  years are compared to each other in
  Fig.~\ref{fig:SpectraAllYears}. The 2003 spectrum is obtained from
  an \emph{ON/OFF} analysis, with the set of special two-telescope
  cuts mentioned above. Note that these cuts were also applied to
  obtain the spectrum shown in Fig.~3 of \citet{Hess1713a}, which
  stops at 10~TeV. Here, however, the 2003 spectrum extends to
  energies beyond 30~TeV. The difference between the two analyses is
  the energy range of simulations used to generate effective gamma-ray
  detection areas (needed for spectral analysis). In the old analysis,
  gamma rays were simulated up to 20~TeV, permitting energy
  reconstruction only up to $\sim10$~TeV (allowing for a maximum
  reconstruction bias of 10\%). Here, in the present analysis,
  simulations up to 100~TeV are available for zenith angles smaller
  than $60\degr$, up to 200~TeV for angles from $60\degr$ to
  $63\degr$, and up to $400$~TeV for zenith angles up to a maximum of
  $70\degr$. Hence the increased energy coverage. Note that good
  agreement is found between the 2003 spectrum shown here and the one
  published previously in \citet{Hess1713a} in the energy range from 1
  to 10~TeV.

  The spectra determined from the 2004 and the 2005 data in
  Fig.~\ref{fig:SpectraAllYears} are obtained with the
  \emph{reflected-region}-background model. Therefore, data where the
  observation position was within the SNR region are disregarded. For
  the purpose of comparison of the different data sets this approach
  seems reasonable, no attempt to analyse the remaining data with an
  \emph{ON/OFF}-background approach is pursued. The corresponding
  event statistics for the spectra shown in the figure are listed in
  Table~\ref{table:std_cuts_stats}.

  In order to compare data recorded in different years, a correction
  for the variation of optical efficiency of the telescope system must
  be applied. The efficiency degrades with time, mostly due to
  degradations of mirror reflectivity. As described in detail in
  \citet{HessCrab}, this worsening of the actual efficiency with
  respect to the simulated one causes a shift in the absolute energy
  scale. This shift can be corrected using measured images of local
  muons, for which the light yield is predictable. Based on the
  prediction and the simulated light yield, an average energy
  correction factor is determined for the data of each of the three
  years separately. The resulting average values are 1.12 for 2003 and
  2004, and 1.30 for 2005. These correction factors are used to
  correct the reconstructed energies thereby enabling direct
  comparisons between different years. Note that a correction
  factor is needed already for the first data set of 2003 since the
  Monte-Carlo simulations refer to new mirrors, but in 2003 the first
  H.E.S.S.\ telescope was already one year old. In 2004, the total
  optical efficiency of the system remained the same because of the
  inclusion of two telescopes with nominal efficiency, thereby
  cancelling the aging effects of the first two telescopes.

  The spectra shown in Fig.~\ref{fig:SpectraAllYears} are after
  correction. Very good agreement is found between the different
  years. The measured spectral shape remains unchanged over time. The
  absolute flux levels are well within the systematic uncertainty of
  20\%. As expected for an object like RX~J1713.7$-$3946, no flux variation is seen
  on yearly timescales. Clearly, the performance of the telescope
  system is under good control, the correction of the optical
  degradation by means of energy correction factors determined from
  ``muon efficiencies'' works reasonably well~(see also
  \citet{HessCrab}). Note that without correction of aging effects,
  flux differences between 2004 and 2005 are on the order 40\%.
  
  \begin{figure}
    \resizebox{\hsize}{!}{\includegraphics[draft=false]{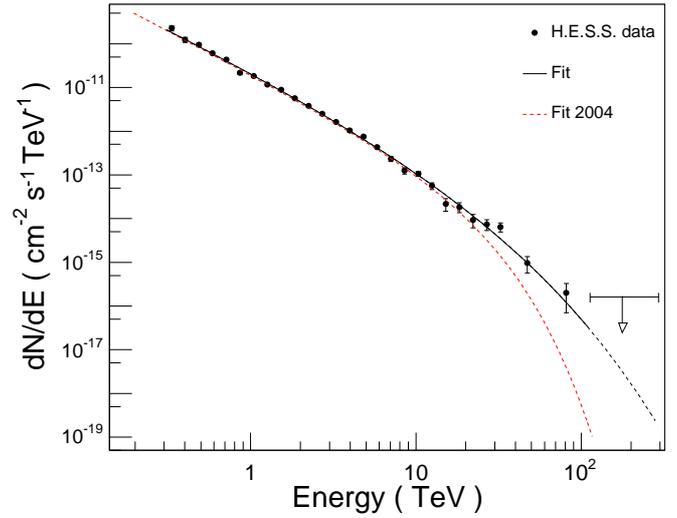}}
    \caption{Combined H.E.S.S.\ gamma-ray spectrum of RX~J1713.7$-$3946\ generated
      from data of 2003, 2004, and 2005 (\emph{Data set III},
      Table~\ref{table:obstimes}). Data are corrected for the
      variation of optical efficiency. Error bars are $\pm 1 \sigma$
      statistical errors. These data might be described by a power law
      with exponential cutoff of the form $dN / dE = I_0 \,
      E^{-\Gamma} \, \exp\left(-(E/E_\mathrm{c})^\beta\right)$. The best
      fit result (black solid line) is given here for $\beta=0.5$
      (fixed), $\Gamma = 1.8$, and $E_\mathrm{c} = 3.7$~TeV (cf.\
      Table~\ref{table:fits} for the exact values). Note that the fit
      function extends as dashed black line beyond the fit range for
      illustration. For comparison, the best fit of a power law with
      exponential cutoff and $\beta=1$, obtained solely from the 2004
      data~\protect\citep{Hess1713b}, is shown as dashed red line. A
      model-independent upper limit, indicated by the black arrow, is
      determined in the energy range from 113 to 300~TeV.}
    \label{fig:CombinedSpectrum}
  \end{figure}
  
  \begin{table*}
    \caption{Fit results for different spectral models. The fit range
    is chosen from 0.3 to 113~TeV. The differential flux normalisation
    $I_0$ is given in units of $10^{-12} \, \mathrm{cm}^{-2} \,
    \mathrm{s}^{-1} \, \mathrm{TeV}^{-1}$. Shown are a power-law model
    (row~1), a power law with an exponential cutoff (row~2,~3,~4; the
    cutoff energy $E_\mathrm{c}$ is given in~TeV), a power law with an
    energy dependent photon index (row~5), and a broken power law (row
    6; in the formula, the parameter $S = 0.6$ describes the sharpness
    of the transition from $\Gamma_1$ to $\Gamma_2$ and it is fixed in
    the fit). Note that when fitting a broken power law to the data,
    some of the fit parameters are highly correlated.}
    \label{table:fits}
    \centering
    \begin{tabular}{|l|llll|l|}
      \hline
      Fit Formula & \multicolumn{4}{c}{Fit Parameters} &
      $\chi^2$~(d.o.f.)\\[1.5ex]
      \hline \hline
      $I_0\ E ^ {-\Gamma}$ & $I_0 = 20.5
      \pm 0.4$ & $\Gamma = 2.32 \pm 0.01$ & & &
      145.6 (25)\\[1.4ex]\hline
      & $I_0 = 21.3
      \pm 0.5$ & $\Gamma = 2.04 \pm 0.04$ & $E_{\mathrm{c}} = 17.9 \pm
      3.3 $ & $\beta = 1.0$ & 39.5 (24)\\[1.4ex]
      $I_0\ E ^ {-\Gamma}\ \exp\left(- ( E / E_{\mathrm{c}})^\beta\right)$ & $I_0 = 34.1
      \pm 2.5$ & $\Gamma = 1.79 \pm 0.06$ & $E_{\mathrm{c}} = 3.7 \pm
      1.0 $ & $\beta = 0.5$ & 34.3 (24)\\[1.4ex]
      & $I_0 = 40.5
      \pm 1.5$ & $\Gamma = 1.74 \pm 0.02$ & $E_{\mathrm{c}} = 2.3 \pm
      0.2 $ & $\beta = 0.45$ & 34.2 (24)\\[1.4ex]\hline
      $I_0\ E ^ {-\Gamma + \, \beta\ \log E}$ & $I_0 = 20.6
      \pm 0.5$ & $\Gamma = 2.02 \pm 0.04$ & $\beta = -0.29 \pm
      0.03 $ & & 38.8 (24)\\[1.4ex]\hline
      $I_0\ \left(E / E_{\mathrm{B}} \right) ^ {-\Gamma_1}\ \left( 1 +
      ( E / E_{\mathrm{B}} ) ^ {1 / S } \right) ^ {\, S \, (\Gamma_1
      - \Gamma_2)}$ & $I_0 = 0.5 \pm 0.4$ & $\Gamma_1
      = 2.00 \pm 0.05$ & $\Gamma_2 = 3.1 \pm 0.2$ & $ E_{\mathrm{B}} =
      6.6 \pm 2.2 $ & 29.8 (23)\\\hline
    \end{tabular}
  \end{table*}

  The combined data of three years are shown in
  Fig.~\ref{fig:CombinedSpectrum}. This energy spectrum of the whole
  SNR region corresponds to $91$~hours of H.E.S.S.\ observations
  (\emph{Data set III}, Table~\ref{table:obstimes}). It is generated
  analysing the 2003 data separately, with the \emph{ON/OFF} approach
  and the two-telescope analysis. The 2004 and 2005 data are analysed
  together, with the \emph{reflected-region} background and the
  nominal 80-photo-electrons cut. As shown in
  Table~\ref{table:obstimes}, a fraction of the data was recorded with
  wobble offsets smaller than $0.7\degr$. For this part, the
  \emph{ON/OFF} method is applied. Average energy-correction factors
  for each of the three subsets of data separately are determined as
  explained above. Having analysed the data separately to obtain
  suitable background estimates for the SNR region, spectra are then
  combined to yield the final spectrum shown in
  Fig.~\ref{fig:CombinedSpectrum}. Systematic checks included the
  application of tighter cuts on the image amplitude to the 2004 and
  2005 data and separate analysis of data recorded under small and
  large zenith angles (below and above $50\degr$). While the spectra
  determined with different cuts are fully compatible, a slight flux
  overestimation is found for the large zenith-angle data, on the 10\%
  level. The investigations of systematic uncertainties at the largest
  zenith angles are still underway, but since the effect on the final
  spectrum is small, $< 5\%$, the combined spectrum given here
  includes all data, up to zenith angles of $70\degr$.
  
  The energy binning of the differential flux shown in the figure is
  chosen to be 12 bins per decade. For the final two points of the
  spectrum, beyond energies of 30~TeV, the binning is three times
  coarser, 4 bins per decade, accounting for decreasing event
  statistics at the highest energies. For the actual positioning of
  the flux points within an energy bin, the method proposed in
  \citet{LaffertyWyatt} is adopted: the point appears at the energy
  value, where the flux value predicted by an effective-area weighted
  model spectral shape (a power law with exponential cutoff) is equal
  to the mean flux value over the energy bin. Note that this is only
  relevant for wide bins. Here, for the spectrum of
  Fig.~\ref{fig:CombinedSpectrum}, the procedure results in flux
  points that are placed within 1\% of the central energy value of the
  bin. Only for the two last points the difference is considerable,
  they end up at 7 and 12\% smaller energy values than the bin centre.

  The combined H.E.S.S.\ spectrum of RX~J1713.7$-$3946\ shown in
  Fig.~\ref{fig:CombinedSpectrum} extends over almost three decades in
  energy, and is compatible with previous H.E.S.S.\
  measurements. \citet{KelnerAharonian} have recently presented a new
  analytical expression (a modified exponential cutoff with exponent
  $\beta$) for secondary gamma-ray spectra from inelastic
  proton-proton interactions based on Monte-Carlo
  simulations. Table~\ref{table:fits} provides the results of fitting
  this function, in addition to several other functional forms to the
  data. A pure power-law model is clearly ruled out, the alternative
  spectral models provide significantly better descriptions of the
  data. For the modified exponential-cutoff shape, the exponent
  $\beta$ was fixed at 1.0, 0.5 and 0.45. The latter $\beta$ value was
  chosen because it yields the smallest $\chi^2$ value. The cutoff
  energy of the gamma-ray spectrum was found to vary depending on
  $\beta$, values of $\sim18$~TeV, $\sim3.7$~TeV, and $\sim2.3$~TeV
  were fit, and none of the three $\beta$ values is statistically
  favoured over the other taking the value of $\chi^2$ as
  measure. Under the assumption that indeed the VHE gamma rays are due
  to cosmic-ray protons interacting with ambient matter and subsequent
  $\pi^0$ decay, one might get an idea of the parameters of the parent
  proton spectrum. Following the approach of \citet{KelnerAharonian},
  proton cutoff values in the range of 50 to 150~TeV with spectral
  indices $\alpha$ ranging from 1.7 to 2.1 would be fully compatible
  with the gamma-ray data presented here. Note, however, that a proton
  cutoff in the 100~TeV range does not mean the spectrum terminates at
  this energy. Especially in case of a hard power-law index $\alpha <
  2$ there would be a sufficient number of protons beyond the cutoff
  energy.

  Combining the data of three years it is possible to extend the
  gamma-ray spectrum up to energies beyond 30~TeV. Taking all events
  with energies above 30~TeV, the cumulative significance is
  $4.8~\sigma$. Table~\ref{table:stats_spectrum_allyears} lists all
  the flux points together with bin-by-bin event statistics.
  
  \begin{table*}
    \caption{Flux points including relevant event statistics are
      listed for the spectrum of the combined H.E.S.S.\ data set, shown
      in Fig.~\ref{fig:CombinedSpectrum}. For all 28 bins, the energy,
      the number of signal and background counts (ON and OFF), the
      normalisation factor $\alpha$, the statistical significance
      $\sigma$, the gamma-ray flux and the energy range of the bin are
      given. The significance is calculated following
      \citet{LiMa}. For the final bin, as it has only marginally
      positive significance, we list both the actual flux point and
      the $2\sigma$ upper limit (which is drawn in
      Fig.~\ref{fig:CombinedSpectrum}). Note that the energy and flux
      values given here are corrected for the variation of optical
      efficiency, as discussed in the main text.}
    \label{table:stats_spectrum_allyears}
    \centering
    \begin{tabular*}{\textwidth}{@{\extracolsep{\fill}}|c|c|c|c|c|c|c|c|}
      \hline
      \# & E ~~ (TeV) & ON & OFF & $\alpha$ & $\sigma$ & Flux ~~ $(\mathrm{cm}^{-2}
      \mathrm{s}^{-1})$ & Range (TeV)\\\hline\hline
      1 & 0.33 & 5890 & 5134 & 1.00 & 7.2 & $(2.73 \pm 0.38) ~~ \times ~~ 10^{-10}$ & 0.30 -- 0.37\\\hline
      2 & 0.40 & 5583 & 4797 & 1.00 & 7.7 & $(1.48 \pm 0.19) ~~ \times ~~ 10^{-10}$ & 0.37 -- 0.44\\\hline
      3 & 0.49 & 4878 & 4010 & 0.97 & 10.5 & $(1.13 \pm 0.11) ~~ \times ~~ 10^{-10}$ & 0.44 -- 0.54\\\hline
      4 & 0.59 & 4202 & 3409 & 0.94 & 11.6 & $(7.22 \pm 0.63) ~~ \times ~~ 10^{-11}$ & 0.54 -- 0.65\\\hline
      5 & 0.71 & 3900 & 2941 & 0.94 & 14.2 & $(5.20 \pm 0.37) ~~ \times ~~ 10^{-11}$ & 0.65 -- 0.79\\\hline
      6 & 0.86 & 3682 & 2833 & 0.97 & 11.9 & $(2.56 \pm 0.22) ~~ \times ~~ 10^{-11}$ & 0.79 -- 0.95\\\hline
      7 & 1.04 & 3881 & 2643 & 0.98 & 16.1 & $(2.17 \pm 0.14) ~~ \times ~~ 10^{-11}$ & 0.95 -- 1.15\\\hline
      8 & 1.26 & 3982 & 2758 & 0.97 & 16.0 & $(1.40 \pm 0.09) ~~ \times ~~ 10^{-11}$ & 1.15 -- 1.39\\\hline
      9 & 1.53 & 4076 & 2661 & 0.98 & 17.9 & $(1.06 \pm 0.06) ~~ \times ~~ 10^{-11}$ & 1.39 -- 1.69\\\hline
      10 & 1.85 & 3873 & 2603 & 0.97 & 17.0 & $(6.71 \pm 0.40) ~~ \times ~~ 10^{-12}$ & 1.69 -- 2.04\\\hline
      11 & 2.24 & 3452 & 2251 & 0.98 & 16.8 & $(4.50 \pm 0.27) ~~ \times ~~ 10^{-12}$ & 2.04 -- 2.47\\\hline
      12 & 2.71 & 3215 & 2113 & 0.98 & 15.9 & $(2.97 \pm 0.19) ~~ \times ~~ 10^{-12}$ & 2.47 -- 2.99\\\hline
      13 & 3.28 & 3075 & 2081 & 0.98 & 14.6 & $(1.95 \pm 0.13) ~~ \times ~~ 10^{-12}$ & 2.99 -- 3.63\\\hline
      14 & 3.98 & 2915 & 2057 & 0.98 & 12.9 & $(1.24 \pm 0.10) ~~ \times ~~ 10^{-12}$ & 3.63 -- 4.39\\\hline
      15 & 4.81 & 2537 & 1721 & 0.98 & 13.1 & $(8.91 \pm 0.68) ~~ \times ~~ 10^{-13}$ & 4.39 -- 5.31\\\hline
      16 & 5.82 & 2183 & 1555 & 0.98 & 10.8 & $(5.18 \pm 0.48) ~~ \times ~~ 10^{-13}$ & 5.31 -- 6.43\\\hline
      17 & 7.05 & 1961 & 1525 & 0.98 & 7.9 & $(2.77 \pm 0.35) ~~ \times ~~ 10^{-13}$ & 6.43 -- 7.79\\\hline
      18 & 8.53 & 1507 & 1208 & 0.98 & 6.2 & $(1.49 \pm 0.24) ~~ \times ~~ 10^{-13}$ & 7.79 -- 9.43\\\hline
      19 & 10.33 & 1211 & 881 & 0.98 & 7.6 & $(1.27 \pm 0.17) ~~ \times ~~ 10^{-13}$ & 9.43 -- 11.41\\\hline
      20 & 12.51 & 881 & 664 & 0.99 & 5.8 & $(6.69 \pm 1.15) ~~ \times ~~ 10^{-14}$ & 11.41 -- 13.81\\\hline
      21 & 15.14 & 652 & 551 & 0.99 & 3.2 & $(2.58 \pm 0.82) ~~ \times ~~ 10^{-14}$ & 13.81 -- 16.72\\\hline
      22 & 18.32 & 473 & 364 & 0.99 & 4.0 & $(2.19 \pm 0.55) ~~ \times ~~ 10^{-14}$ & 16.72 -- 20.24\\\hline
      23 & 22.18 & 327 & 260 & 0.99 & 2.9 & $(1.10 \pm 0.38) ~~ \times ~~ 10^{-14}$ & 20.24 -- 24.50\\\hline
      24 & 26.85 & 220 & 153 & 0.99 & 3.6 & $(8.82 \pm 2.46) ~~ \times ~~ 10^{-15}$ & 24.50 -- 29.66\\\hline
      25 & 32.50 & 182 & 110 & 0.99 & 4.3 & $(7.70 \pm 1.79) ~~ \times ~~ 10^{-15}$ & 29.66 -- 35.91\\\hline
      26 & 47.19 & 227 & 180 & 0.99 & 2.5 & $(1.15 \pm 0.47) ~~ \times ~~ 10^{-15}$ & 35.91 -- 63.71\\\hline
      27 & 81.26 & 51 & 37 & 0.99 & 1.5 & $(2.36 \pm 1.55) ~~ \times ~~ 10^{-16}$ & 63.71 -- 113.02\\\hline
      &&&&& 0.6 & $\left(3.77^{+6.39}_{-3.77}\right) ~~ \times ~~ 10^{-17}$ & \\[1.5ex]
      \raisebox{2.25ex}[-2.25ex]{28} & \raisebox{2.25ex}[-2.25ex]{169.79} & \raisebox{2.25ex}[-2.25ex]{14} & \raisebox{2.25ex}[-2.25ex]{11} & \raisebox{2.25ex}[-2.25ex]{1.00} & Upper Limit & $1.6 \times 10^{-16}$ & \raisebox{2.25ex}[-2.25ex]{113.02 -- 293.82}\\\hline
    \end{tabular*}
  \end{table*}

  \section{Summary \& Discussion}
  \label{sec:summary}

  The complete H.E.S.S.\ data set of the SNR RX~J1713.7$-$3946\ recorded from 2003 to
  2005 is presented here. When analysing the data of different years
  separately and comparing them to each other, a very good agreement
  is found for both the gamma-ray morphology and the differential
  energy spectra. The H.E.S.S.\ telescope system obviously operates
  stably over the course of three years, if one takes known aging
  effects into account.

  A combined gamma-ray image using $\sim63$~hours of H.E.S.S.\
  observations in 2004 and 2005 was generated achieving an
  unprecedented angular resolution of $0.06\degr$. The morphology of
  RX~J1713.7$-$3946\ in VHE gamma rays confirms its earlier
  characterisation~\citep{Hess1713b} of a thick, almost circular shell
  structure with the brightest regions in the northwest, very similar
  to the X-ray image of this source. The gamma-ray spectrum of the
  combined H.E.S.S.\ data over three years on RX~J1713.7$-$3946\ extends over three
  orders of magnitude in energy. Although at the edge of sufficient
  statistical significance, the high-energy end of the gamma-ray
  spectrum approaches 100~TeV with significant emission $(4.8\sigma)$
  beyond 30~TeV. Given the systematic uncertainties in the spectral
  determination at these highest energies and comparable statistical
  uncertainties despite the long exposure time, this measurement is
  presumably close to what can be studied with the current generation
  of imaging atmospheric Cherenkov telescopes.

  From the largest measured gamma-ray energies one can estimate the
  corresponding energy of the primary particles. If VHE gamma rays are
  produced via $\pi^0$ decay following inelastic proton-proton
  interactions, gamma-ray energies of 30~TeV imply that primary
  protons are accelerated to $30~\mathrm{TeV} / 0.15 = 200~\mathrm{TeV}$ in
  the shell of RX~J1713.7$-$3946. On the other hand, if the gamma rays are due to
  Inverse Compton scattering of VHE electrons, accelerated in the
  shell, off Cosmic--Microwave--Background photons (neglecting the
  presumably small contributions from starlight and infrared photons),
  the electron energies at the current epoch can be estimated in the
  Thompson regime as $E_\mathrm{e}~\approx~20~\sqrt{E_\gamma}~\mathrm{TeV}
  \approx 110~\mathrm{TeV}$. At these large energies Klein--Nishina
  effects start to be important and reduce the maximum energy slightly
  such that $\sim100~\mathrm{TeV}$ is a realistic estimate.

  If one considers the functional representations found for the fit of
  the gamma-ray spectrum of RX~J1713.7$-$3946\ (c.f.\ Table~\ref{table:fits}), the
  basic findings of~\citet{Hess1713b} are confirmed with improved
  statistics and increased energy coverage: a pure power-law spectral
  shape is clearly ruled out, alternative models like a broken
  power-law, a power with energy-dependent exponent, and a power law
  with exponential cutoff describe the data significantly
  better. Assuming an exponential--cutoff shape, a ``slow'' cutoff
  with exponent $\beta=0.5$, as suggested by detailed Monte-Carlo
  simulations~\citep{KelnerAharonian}, yields a perfect description of
  the data, however, different values of $\beta$ cannot be
  distinguished, but would rather require better event statistics at
  the highest energies.

  Given the good agreement of the results presented here with the
  previously published ones, our restrictive conclusions regarding the
  nature of the parent particles remain unchanged to those outlined in
  ~\citet{Hess1713b}. Both scenarios with a leptonic, or hadronic
  primary particle distribution are able to accommodate an
  exponential--cutoff shape with an index of $\beta$ $\sim
  0.5$. However, if the mean magnetic field in the SNR region is
  indeed strongly amplified by the shock to values well beyond typical
  interstellar fields, the hadronic nature of the observed gamma-ray
  emission would be difficult to conceal and this latter emission
  scenario would be clearly favoured~\citep{BerezhkoVoelk}.

  With the deep H.E.S.S.\ observations of RX~J1713.7$-$3946\ we approach now energies,
  at which attenuation due to pair production on the Galactic
  interstellar radiation field begins to affect the gamma-ray
  spectrum~\citep{2006A&A...449..641Z}. At the currently measured
  maximum energy this effect is negligible, particularly since RX~J1713.7$-$3946\
  is neither in the direction of the Galactic Center (more than
  $12\degr$ angular separation in line-of-sight), nor is it at the
  distance where the interstellar radiation field
  peaks~\citep{2006ApJ...640L.155M}. RX~J1713.7$-$3946\ will therefore presumably
  not be the astronomical source, where we will obtain a clear
  observational confirmation of the attenuation of gamma rays due to
  the interstellar radiation field. However, RX~J1713.7$-$3946\ remains an
  exceptional SNR in respect of its VHE gamma-ray observability, being
  at present the remnant with the widest possible coverage along the
  electromagnetic spectrum. The H.E.S.S.\ measurement of significant
  gamma-ray emission beyond 30~TeV without indication of a termination
  of the high-energy spectrum provides proof of particle acceleration
  in the shell of RX~J1713.7$-$3946\ beyond $10^{14}$~eV, up to energies which
  start to approach the region of the cosmic-ray \emph{knee}.

  \section*{Acknowledgments}
  The support of the Namibian authorities and of the University of
  Namibia in facilitating the construction and operation of
  H.E.S.S. is gratefully acknowledged, as is the support by the German
  Ministry for Education and Research (BMBF), the Max Planck Society,
  the French Ministry for Research, the CNRS-IN2P3 and the
  Astroparticle Interdisciplinary Programme of the CNRS, the
  U.K. Particle Physics and Astronomy Research Council (PPARC), the
  IPNP of the Charles University, the South African Department of
  Science and Technology and National Research Foundation, and by the
  University of Namibia. We appreciate the excellent work of the
  technical support staff in Berlin, Durham, Hamburg, Heidelberg,
  Palaiseau, Paris, Saclay, and in Namibia in the construction and
  operation of the equipment.

  \bibliographystyle{aa}
  \bibliography{main}

\begin{thebibliography}{25}
\expandafter\ifx\csname natexlab\endcsname\relax\def\natexlab#1{#1}\fi

\bibitem[{{Aharonian} {et~al.}(2004{\natexlab{a}})}]{HessCalib}
{Aharonian et al. \textit{(H.E.S.S. Collaboration)}} 2004{\natexlab{a}}, APh, 22, 109

\bibitem[{{Aharonian} {et~al.}(2004{\natexlab{b}})}]{Hess1713a}
{Aharonian et al. \textit{(H.E.S.S. Collaboration)}} 2004{\natexlab{b}}, Nature, 432,
  75

\bibitem[{{Aharonian} {et~al.}(2005{\natexlab{a}})}]{HessVelaJr_a}
{Aharonian et al. \textit{(H.E.S.S. Collaboration)}} 2005{\natexlab{a}}, \aap, 437, L7

\bibitem[{{Aharonian} {et~al.}(2005{\natexlab{b}})}]{Hess2155}
{Aharonian et al. \textit{(H.E.S.S. Collaboration)}} 2005{\natexlab{b}}, A\&A, 430,
  865

\bibitem[{{Aharonian} {et~al.}(2006{\natexlab{a}})}]{HessVelaJr_b}
{Aharonian et al. \textit{(H.E.S.S. Collaboration)}} 2006{\natexlab{a}}, submitted to
  \apj

\bibitem[{{Aharonian} {et~al.}(2006{\natexlab{b}})}]{Hess1713b}
{Aharonian et al. \textit{(H.E.S.S. Collaboration)}} 2006{\natexlab{b}}, \aap, 449,
  223

\bibitem[{{Aharonian} {et~al.}(2006{\natexlab{c}})}]{HessCrab}
{Aharonian et al. \textit{(H.E.S.S. Collaboration)}} 2006{\natexlab{c}}, \aap, 457,
  899

\bibitem[{{Berezhko} \& {V{\"o}lk}(2006)}]{BerezhkoVoelk}
{Berezhko}, E.~G. \& {V{\"o}lk}, H.~J. 2006, \aap, 451, 981

\bibitem[{{Berge}(2006)}]{PhDBerge}
{Berge}, D. 2006, PhD thesis, Ruprecht-Karls Universit\"at, Heidelberg,
  http://www.ub.uni-heidelberg.de/archiv/6156

\bibitem[{{Bernl{\"o}hr} {et~al.}(2003){Bernl{\"o}hr}, {Carrol}, {Cornils},
  {Elfahem}, {Espigat}, {Gillessen}, {Heinzelmann}, {Hermann}, {Hofmann},
  {Horns}, {Jung}, {Kankanyan}, {Katona}, {Khelifi}, {Krawczynski}, {Panter},
  {Punch}, {Rayner}, {Rowell}, {Tluczykont}, \& {van Staa}}]{Konrad2003}
{Bernl{\"o}hr}, K., {Carrol}, O., {Cornils}, R., {et~al.} 2003, Astroparticle
  Physics, 20, 111

\bibitem[{{Cornils} {et~al.}(2003){Cornils}, {Gillessen}, {Jung}, {Hofmann},
  {Beilicke}, {Bernl{\" o}hr}, {Carrol}, {Elfahem}, {Heinzelmann}, {Hermann},
  {Horns}, {Kankanyan}, {Katona}, {Krawczynski}, {Panter}, {Rayner}, {Rowell},
  {Tluczykont}, \& {van Staa}}]{CornilsII}
{Cornils}, R., {Gillessen}, S., {Jung}, I., {et~al.} 2003, APh, 20, 129

\bibitem[{{Drury} {et~al.}(1994){Drury}, {Aharonian}, \& {Voelk}}]{DAV}
{Drury}, L.~O., {Aharonian}, F.~A., \& {Voelk}, H.~J. 1994, \aap, 287, 959

\bibitem[{{Funk} {et~al.}(2004){Funk}, {Hermann}, {Hinton}, {Berge},
  {Bernl{\"o}hr}, {Hofmann}, {Nayman}, {Toussenel}, \&
  {Vincent}}]{FunkTriggerPaper}
{Funk}, S., {Hermann}, G., {Hinton}, J., {et~al.} 2004, Astroparticle Physics,
  22, 285

\bibitem[{{Hillas}(2005)}]{HillasReview}
{Hillas}, A.~M. 2005, Journal of Physics G Nuclear Physics, 31, R95

\bibitem[{{Hinton} {et~al.}(2005){Hinton}, {Berge}, \& {Funk}}]{BackgroundProc}
{Hinton}, J., {Berge}, D., \& {Funk}, S. 2005, in Conference Proceedings
  "Towards a Network of Atmospheric Cherenkov Detectors VII", Palaiseau, France

\bibitem[{{Hiraga} {et~al.}(2005){Hiraga}, {Uchiyama}, {Takahashi}, \&
  {Aharonian}}]{HiragaXMM}
{Hiraga}, J.~S., {Uchiyama}, Y., {Takahashi}, T., \& {Aharonian}, F.~A. 2005,
  A\&A, 431, 953

\bibitem[{{Hofmann} {et~al.}(1999){Hofmann}, {Jung}, {Konopelko},
  {Krawczynski}, {Lampeitl}, \& {P{\"u}hlhofer}}]{HofmannShowerReco}
{Hofmann}, W., {Jung}, I., {Konopelko}, A., {et~al.} 1999, Astroparticle
  Physics, 12, 135

\bibitem[{{Hofmann}(2005)}]{HofmannStatusHess}
{Hofmann, W. \textit{(H.E.S.S. Collaboration)}} 2005, in Conference Proceedings
  "Towards a Network of Atmospheric Cherenkov Detectors VII", Palaiseau,
  France, 43--56

\bibitem[{{Kelner} {et~al.}(2006){Kelner}, {Aharonian}, \&
  {Bugayov}}]{KelnerAharonian}
{Kelner}, S.~R., {Aharonian}, F.~A., \& {Bugayov}, V.~V. 2006, \prd, 74, 034018

\bibitem[{{Lafferty} \& {Wyatt}(1995)}]{LaffertyWyatt}
{Lafferty}, G.~D. \& {Wyatt}, T.~R. 1995, Nuclear Instruments and Methods in
  Physics Research A, 355, 541

\bibitem[{{Li} \& {Ma}(1983)}]{LiMa}
{Li}, T.-P. \& {Ma}, Y.-Q. 1983, Astrophysical Journal, 272, 317

\bibitem[{{Moskalenko} {et~al.}(2006){Moskalenko}, {Porter}, \&
  {Strong}}]{2006ApJ...640L.155M}
{Moskalenko}, I.~V., {Porter}, T.~A., \& {Strong}, A.~W. 2006, \apjl, 640, L155

\bibitem[{{Uchiyama} {et~al.}(2002){Uchiyama}, {Takahashi}, \&
  {Aharonian}}]{UchiyamaAsca}
{Uchiyama}, Y., {Takahashi}, T., \& {Aharonian}, F.~A. 2002, PASJ, 54, L73

\bibitem[{{Vincent et al.}(2003)}]{PascalHessCamera}
{Vincent, P. et al. \textit{(H.E.S.S. Collaboration)}} 2003, in Proc. 28th ICRC, 2887

\bibitem[{{Zhang} {et~al.}(2006){Zhang}, {Bi}, \& {Hu}}]{2006A&A...449..641Z}
{Zhang}, J.-L., {Bi}, X.-J., \& {Hu}, H.-B. 2006, \aap, 449, 641

\end{thebibliography}
\end{document}